\documentclass[%
 reprint,
 nofootinbib,
 amsmath,amssymb,
 aps,
 prl,
]{revtex4-1}
\usepackage{amsmath}
\usepackage{hyperref}
\usepackage{cleveref}
\usepackage{mathtools}
\usepackage{graphicx}% Include figure files
\usepackage{dcolumn}% Align table columns on decimal point
\usepackage{bm}% bold math
\usepackage{enumitem,kantlipsum}
\newcommand{\bb}{\hat{b}}

\begin{document}

%\preprint{APS/123-QED}

\title{A Quantum Spectrometer for Arbitrary Noise}% Force line breaks with \\
% \thanks{A footnote to the article title}%

\author{Daniel Goldwater}
%   \affiliation{Department of Mathematical Sciences, University of Nottingham,\\  Nottingham NG7 2QL, United Kingdom}
  %Lines break automatically or can be forced with \\
\author{Peter Barker}%
%  \email{dangoldwater@gmail.com}
\affiliation{%
 Department of Physics and Astronomy, University College London, Gower Street, London WC1E 6BT, United Kingdom
}%

% \collaboration{MUSO Collaboration}%\noaffiliation

\author{Angelo Bassi}
%  \homepage{http://www.Second.institution.edu/~Charlie.Author}
\affiliation{Department of Physics, University of Trieste, Strada Costiera 11, 34151 Trieste, }
 \affiliation{Istituto Nazionale di Fisica Nucleare, Trieste Section, Via Valerio 2, 34127 Trieste, Italy}%
% \collaboration{CLEO Collaboration}%\noaffiliation
\author{Sandro Donadi}
\affiliation{Frankfurt Institute for Advanced Studies (FIAS), Ruth-Moufang-Stra{\ss}e 1, 60438 Frankfurt am Main, Germany}
% Institut f\"ur Theoretische Physik, Universit\"at Ulm, D-89069, Germany}
\date{\today}% It is always \today, today,
             %  but any date may be explicitly specified

\begin{abstract}
We present a technique for recovering the spectrum of a non-Markovian bosonic bath and/or non-Markovian noises coupled to an harmonic oscillator. The treatment is valid under the conditions that the environment is large and hot compared to the oscillator, and that its temporal auto-correlation functions are symmetric with respect to time translation and reflection -- criteria which we consider fairly minimal. We model a demonstration of the technique as deployed in the experimental scenario of a nanosphere levitated in a Paul trap, and show that it would effectively probe the spectrum of an electric field noise source from $10^2\,-\,10^6$ Hz with a resolution inversely proportional to the measurement time. This technique may be deployed in quantum sensing, metrology, computing, and in experimental probes of foundational questions.
\end{abstract}

\pacs{Valid PACS appear here}% PACS, the Physics and Astronomy
                             % Classification Scheme.
%\keywords{Suggested keywords}%Use showkeys class option if keyword
                              %display desired
\maketitle

\section{\label{sec.Introduction} Introduction}
Noise is an unavoidable feature of all physical measurements, and often their main impediment. The study of noise in quantum physics is a field in its own right \cite{QuantumNoise,Clerk2010}, and the peculiar structure of the noise spectrum is of central importance to quantum metrology \cite{giovannetti2006quantum} and more generally to the theory of open quantum systems \cite{breuer2016colloquium}. %It has been shown \cite{Chin2012} that, far from precluding quantum sensitivity, the non-Markovianity of the quantum noise in metrological scenarios is essential to the success of the protocol.
However, in the majority of treatments, the exact nature of the noise spectrum is left as an unknown quantity --  if it not idealised as Markovian, it will typically be ascribed a simple structure, such as Ohmic \cite{Chin2012}, though a larger taxonomy exists \cite{rivas2014quantum}. 

In this article we introduce a scheme which is capable of measuring the spectrum of arbitrary noises or bosonic baths. In itself this constitutes a new quantum sensing tool comparable to other spectrometers \cite{frey2017application,norris2018optimally,Silva2017}, but it may also be used to improve existing techniques by helping metrological schemes which deal with noise \cite{Giovannetti2011}. It can be enacted on any harmonic oscillator which can be prepared in a cold initial state, and whose resonant frequency may be varied. Since the vast majority of quantum sensors, and indeed quantum experiments, are based around the behaviour of harmonic oscillators, this implies a potentially wide range of applications. 

One direct utilisation which we will demonstrate in this work would be to study the structure of the electric field noise found in electric particle traps, which arises from the electrodes and is determined by several factors \cite{safavi2013influence} in a way not yet fully understood \cite{brownnutt2015ion,safavi2011microscopic}. To make this demonstration, we will model the specific scenario of an electrically levitated charged nanosphere; in such a setting, the environmental noise conditions are such that an accurate reconstruction of the electric field noise spectrum ought to be possible over a wide range - between $10^2$ Hz and $10^6$ Hz. Further, this ability to accurately characterise the environmental noise spectrum may find application in quantum computing, where such knowledge would enable the development of optimised dynamical decoupling protocols tailored to the specific environment of the qubit(s)~\cite{viola1999dynamical}. It may also find applications in short-range force sensing \cite{hempston2017force,winstone2017direct}, where the spectrum characterising the interaction between the force being studied and the behaviour of the oscillator can be subjected to a similar treatment.

Here we make a proposal in three parts. First, we describe a mathematical formalism through which the spectrum of a general bosonic bath coupled to a quantum harmonic oscillator may be recovered through experiment (granted certain assumptions, and up to a degree of uncertainty). Second, we propose a specific experimental scenario which is particularly suited to this task (that of a levitated nanoparticle) and demonstrate its theoretical performance. Third, we examine an example case -- that of an electric field noise with a non-trivial spectrum -- and show how the experimental scenario of the previous section would be able to recover this spectrum through the suggested technique. 
% Finally, we show how the technique might be utilised towards a foundational purpose: testing and characterising models of spontaneous wavefunction collapse in which the putative noise field has an arbitrary spectrum \cite{adler2007collapse,adler2008collapse,Carlesso2018color,Ferialdi2012a}. 
% Finally, we show how the technique presented can be used for testing collapse models with non-white noises. 

% it remains standard to treat the any form of quantum noise as Markovian, possessing a `white' spectrum. Of course, this assumption is often deeply unphysical. No noise source can, in truth, possess a truly white spectrum, as this would correspond to its originating processes occurring both infinitely fast and infinitely slowly. 

% \section{\label{sec.Formalism} Formalism}
{\it Formalism --} We  begin by considering a  Hamiltonian which describes an harmonic oscillator $\mathcal{S}$ coupled to a bosonic bath of independent harmonic oscillators $\mathcal{B}$:
\begin{align}
    \hat{H} &= \hat{H}_\mathcal{S}+\hat{H}_\mathcal{B}+\hat{H}_\mathcal{I},\\
    \shortintertext{in which the three terms above represent the system, bath and interaction Hamiltonians respectively. Setting $\hbar=1$, they are given by}
    \hat{H}_\mathcal{S} & = \omega_m  (\hat{a}^\dagger \hat{a}+\frac{1}{2})\label{hs},\\
    \hat{H}_\mathcal{B} &= \sum_\alpha  \omega_\alpha (\bb^\dagger_\alpha \bb_\alpha +\frac{1}{2})\label{hb},\\
    \hat{H}_\mathcal{I} & =- \hat{q} \sum_\alpha g_\alpha\hat{q}_\alpha, \label{hi} 
\end{align} 
where $\omega_m$ represents the mechanical frequency of the oscillator and $\hat{a}^\dagger,\,\hat{a},\,\hat{q}$ give its creation, annihilation and position operators respectively. Similarly, we decompose $\mathcal{B}$ into $\alpha$ modes, whose creation, annihilation and position operators are $\hat{b}_\alpha^\dagger$, $\hat{b}_\alpha$ and $\hat{q}_\alpha$ respectively. The interaction between the system and the $\alpha$ mode of the bath is described through a position-position coupling whose strength is given by $g_\alpha=m_{\alpha}\omega_{\alpha}^{2}$ whose value can be settled by an appropriate choice of the bath oscillator masses. 

The first description of the dynamics of this type of system was given by Caldeira and Leggett in their seminal paper \cite{caldeira1983path}, in which they derived a master equation for this scenario using the Born-Markov approximation, valid in the limit of high temperatures of the bath. This result was improved by Hu, Paz and Zhang \cite{Zhang1992}, who derived a master equation which is exact, and valid for any temperature. The subject has more been recently studied in \cite{ferialdi2017dissipation,carlesso2017adjoint}, where the Hu, Paz and Zhang master equation was derived in the form%\footnotemark[1]{}
\begin{align}\label{HPZ}
\frac{d}{dt}\rho_{t}=&-i[\hat{H}_{S}-\Xi(t)q^{2},\rho_{t}]+\Gamma(t)[\hat{q},[\hat{q},\rho_{t}]]\nonumber\\
&+\Theta(t)[\hat{q},[\hat{p},\rho_{t}]]+i\Upsilon(t)[\hat{q},\{\hat{p},\rho_{t}\}],
\end{align}
with $\rho_t$ the density matrix at time $t$, and $\hat{p}$ the momentum operator. It is eq.\eqref{HPZ} which we shall use as a jumping off point for developing our formalism -- it is very general -- for example, the bath need not be thermal for the equation to be valid. It must, however, begin in a Gaussian state\footnotemark{}. In appendix 1 of the supplementary information we detail an alternative derivation.
\footnotetext{For an example of a spectrometer which could probe non-Gaussian noise, see \cite{sung2019non,Norris2016}.}
%in exchange for a weak coupling condition.
%\footnotetext[2]{Note that in \cite{ferialdi2017dissipation}, $g_\alpha=m_\alpha \omega^2_\alpha$.}

The exact definitions of the time-dependent coefficients $\Xi(t)$, $\Gamma(t)$, $\Theta(t)$ and $\Upsilon(t)$ are given through recursive series expansions and can be found in \cite{ferialdi2017dissipation}. However, in the limit of a weak coupling between the system and bath (a limit which we will now assume), one can safely make a first order approximation which greatly simplifies the expressions for these coefficients. Further to this, we will consider only the regime where the damping effects of the bath upon the system will be negligible compared to its heating effects, which is mathematically equivalent to assuming that the bath correlation function is real \cite{ferialdi2017dissipation}. Note that, in the case of thermal baths, the assumption that dissipative effects are negligible is equivalent to assuming that the temperature of the bath is much higher than that of the system. We will go into further details regarding these simplifications in appendix 1 of the supplementary information, which also includes refs .

Taking these simplifications, one gets  $\Xi(t)=\Upsilon(t)=0$ and 
\begin{align}\label{Gamma}
\Gamma(t)=-\int_{0}^{t}ds\,C(t,s)\cos[\omega_{m}(t-s)],
\end{align}
\begin{align}\label{Theta}
\Theta(t)=\int_{0}^{t}ds\,C(t,s)\frac{\sin[\omega_{m}(t-s)]}{m\omega_{m}},
\end{align}
in which $C(t,s) = {\rm Tr}[\hat{B}(t)\hat{B}(s)\rho_B]$ is the two-time correlation function for the bath operator $\hat{B}=\sum\limits_\alpha g_\alpha \hat{q}_\alpha$. The Fourier transform of $C(t,s)$ -- the spectrum of the noise function in frequency space -- is the object which our spectrometer will ultimately uncover through a study of its impact upon the system. 
%  the second approximation here sounds as though we are assuming the bath to be of infinite temperature, this is actually not the case since we are still dealing with a bath general enough that it might not be thermal.}% CONSIDER IF ADD SOME EQUATIONS WHERE THE NOISE CORRELATION IS WRITTEN IN TERMS OF SPECTRAL DENSITIES. GIVES A CONNECTION WITH TEMPERATURE. IS TO HAVE SOME REASON TO SET THE OTHER TWO CONSTANT =0. 
Using these simplifications, eq. \eqref{HPZ} becomes
\begin{align}\label{HPZnodis}
\frac{d}{dt}\rho_{t}=-i[\hat{H}_{S},\rho_{t}]+\Gamma(t)[\hat{q},[\hat{q},\rho_{t}]]+\Theta(t)[\hat{q},[\hat{p},\rho_{t}]].
\end{align}

For the spectrometer to function as a viable measurement instrument, we must of course select an observable to monitor. Here we show that the number operator $\hat{n} =\hat{a}^\dagger \hat{a}$ is ideal. It is experimentally straightforward to measure, and can be used to recover $C(t,s)$ unambiguously.

The equation of motion for the expected occupation number can be found via 
$
(d/dt)\langle \hat{n}\rangle_t  ={\rm Tr}[\hat{n}\mathcal{L}\rho_t]
$,
in which $\mathcal{L} \rho_t$ is a super-operator on $\rho_t$ which summarizes the right-hand side of eq. \eqref{HPZnodis}. Using this, the cyclicity of the trace, and a little algebra we find that
\begin{align}\label{Eqforn}
\frac{d}{dt}\langle \hat{n}\rangle_{t}=
% -\frac{\Gamma(t)}{2m\omega_{m}}=
\frac{1}{2m\omega_{m}}\int_{0}^{t}C(t,s)\cos[\omega_{m}(t-s)].
\end{align}
Now, if we assume that the correlation function is invariant with respect to both time reversal and time translation, i.e. that $C(t,s)=C(|t-s|)$, we can rewrite the right hand side of eq. \eqref{Eqforn} in a more convenient form by  using the relation
\begin{equation}
    \int_0^t ds\,C(|s-t|)\cos[\omega_m (s-t)]=\frac{1}{2}\int_{-t}^t dy\,C(y)e^{i\omega_m y}.
\end{equation}
Introducing the Fourier expansion of $C(y)$
\begin{align}
C(y)&=\frac{1}{2\pi}\int_{-\infty}^\infty d\nu\,\tilde{C}(\nu)e^{-i\nu y},
\end{align}
eq. \eqref{Eqforn} becomes 
\begin{align}\label{eqforn}
\frac{d}{dt}\langle \hat{n}\rangle_{t}&
% =\frac{1}{8\pi m\omega_{m}}\int_{-t}^{t}dy\,\int_{-\infty}^{\infty}d\nu\,\tilde{C}(\nu)e^{i(\omega_{m}-\nu)y}\nonumber\\
&=\frac{1}{4\pi m\omega_{m}}\int_{-\infty}^{\infty}d\nu\,\tilde{C}(\nu)\frac{\sin[(\omega_{m}-\nu)t]}{(\omega_{m}-\nu)},
\end{align}
which can be solved to get:
\begin{align}
\langle \hat{n}\rangle_{t}=\langle \hat{n}\rangle_{0}+\frac{1}{2\pi m\omega_{m}}\int_{-\infty}^{\infty}d\nu\,\tilde{C}(\nu)\frac{\sin^{2}[(\omega_{m}-\nu)t/2]}{(\omega_{m}-\nu)^{2}}.\label{eq.SpectrIdeal}
\end{align}

    This equation contains the basic capabilities of the spectrometer, and as such it demands some examination. In the white noise limit $\tilde{C}(\nu)=D_p$ with $D_p$ being some positive constant, it gives the well known behavior: $\langle \hat{n}\rangle_{t}=\langle \hat{n}\rangle_{0}+ D_p' t$, where $D_p'=D_p/4m\omega_m$ quantifies the heating rate due to momentum diffusion. Whilst the integral in \eqref{eq.SpectrIdeal} is of course unsolvable without knowing $\tilde{C}(\nu)$, its form allows us to estimate the function of interest. The function $\sin^2(\theta t/2)/\theta^2$ forms a peak around $\theta=0$ of width $4\pi/t$. For us, this means that the envelope of the peak will serve analogously to a flared $\delta$ function, selecting the effects of $\tilde{C}(\nu)$ in the region of $\nu=[ \omega_m-\frac{2\pi}{t},\omega_m+\frac{2\pi}{t}]$. We can deploy this approximation, in eq. (\ref{eq.SpectrIdeal}), replacing the integral with $\tilde{C}(\omega_m)\frac{\pi t}{2}$ and re-arranging to get: 
\begin{equation}
         \tilde{C}(\omega_m)=\frac{4 m \omega_m}{t}\left(\langle n\rangle_t-\langle n\rangle_0\right). \label{eq.SpecReconst1}
\end{equation}
% Using this, we can approximate equation \ref{eq.SpectrIdeal} as 
% \begin{align}
%     \langle n\rangle_t &\simeq \langle n\rangle_0+\frac{1}{2\pi m\omega_m}\frac{\pi t}{2}\tilde{C}(\omega_m)\\
%     \shortintertext{which is easily re-arranged to give}
%     \tilde{C}(\omega_m)&=\frac{4 m \omega_m}{\pi t}\left(\langle n\rangle_t-\langle n\rangle_0\right). 
% \end{align}
Note that although this is appears to be linear in $\omega_m$, the dominant term will typically be the $\langle n\rangle_t$ of eq. \eqref{eq.SpectrIdeal}, in which the coefficient in front of the integral is $\propto1/\omega_m$, neutralising the linear scaling -- and that $\langle n\rangle_t$ has a more complicated dependence upon $\omega_m$ in the integrand related to the noise spectrum.
If $\omega_m$ is changeable, this will allow us to scan through the range of available frequencies and see how $\tilde{C}(\nu)$ behaves across the range with an error $\propto 1/t$. This, then,  forms the protocol for the spectrometer: to take measurements of $\langle \hat{n}\rangle_{t}$ at different values of $\omega_m$, and use the results to reconstruct $\tilde{C}(\nu)$ via eq. \eqref{eq.SpecReconst1}. The accuracy of this reconstruction will be improved with an increasing time $t$ taken per measurement. 

Eq. (\ref{eq.SpectrIdeal}) has the same structure as eq. (18) from \cite{cywinski2008enhance}, which focuses on the decoherence of superconducting qubits resulting from their interaction with a classical noise. However, there are fundamental differences between the two approaches. First of all, the systems interacting with the noise considered in \cite{cywinski2008enhance} are qubits, whilst we consider a harmonic oscillator. More importantly, in the approach used in \cite{cywinski2008enhance}, in order to study the noise spectrum for high frequencies it is necessary to introduce pulse sequences which, if properly designed, allow for the suppression of the effects of the low frequency parts of the spectrum. In this respect, our approach is simpler, since in order to measure different frequency regions of the noise spectrum, one needs simply to change the trap's frequency. 

{\it Experimental Scheme --}
As we can see in eq.~(\ref{eq.SpectrIdeal}), the rate by which the harmonic oscillator will heat depends strongly upon its mechanical frequency $\omega_m$ and the strength of the noise source in the region around that frequency. By taking account of the initial phonon number $n_0$ and the effects of other baths coupled to the system, this can be used to probe the spectrum of the noise  induced upon the system by the bath of interest $\mathcal{B}$. The effectiveness of such a probe will be determined by the following factors:
\begin{enumerate}[wide, labelwidth=!, labelindent=0pt]
    \item[-] The range over which its mechanical frequency $\omega_m$ can be adjusted.
    \item[-] The lower limit of the temperature in which the oscillator can be prepared, with two-fold purpose. Firstly, the colder the oscillator, the more accurate is the approximation made above in which the bath has significantly higher energy than the oscillator, allowing us to ignore damping effects; and secondly, the lower the value of $\langle \hat{n}_0 \rangle$, the lower its spread, and the more accurately $\Delta \hat{n}$ can be deduced.
    \item[-] The accuracy with which the other baths not being measured can be estimated. In particular, the approximation that they are Markovian (and hence possess a flat spectrum) must be reasonable.
    % ; else, any novel frequency dependant features uncovered in the experiment would be ascribed to the wrong bath (unless, of course, this second bath has a coupling to the system which is considerably weaker than that of the bath of interest). It \emph{would} be possible to distinguish between the spectra of more than one non-Markovian bath coupled to a single system, but this would require an extension to the technique described here. 
    \item[-] The accuracy with which $\langle \hat{n} \rangle_t$ can be measured. 
    \item[-] The ability to increase or decrease the coupling to the bath of interest $\mathcal{B}$. This is not strictly necessary, but as shown in \cite{Goldwater2016}, such a capability is essential in certain contexts for distinguishing between the effects of interest and the effects of other baths. 
    % For example, we could in principle distinguish between the effects of electric field noise and gas collision noise affecting a trapped particle by varying the charge and radius of the levitated particle and see how the heating rate varies, since these two noise baths would couple differently via these parameters.
    \item[-] The duration of the experiment -- as shown in the previous section, the longer we can let the system get heated by the bath, the higher the accuracy of the spectrum measurement.    
\end{enumerate}

We claim that an experiment built around a charged levitated nanosphere is well suited to these needs. A `hybrid' type trap, composed of a quadrupole electric field trap working in conjunction with an optical trap as described in \cite{Millen2015,Goldwater2016}, is ideal: it has an exceptionally low noise floor, it is capable of cooling the particle to a very low occupation (in principle to the ground state) using the techniques outlined in \cite{Goldwater2018}, and the mechanical frequency can be changed at will over the impressively wide range from $\sim 100$ Hz to $\sim 10^6$ Hz. The Paul trap used to levitate the nanosphere is structurally the same as those used as ion traps: an oscillating electric field holds a charged particle in an harmonic well with variable frequency. 

The theoretical modelling of noise sources affecting the levitated sphere is a well trodden path \cite{Chang2010,Goldwater2016,rodenburg2016quantum,Romero-Isart2011,Barker2010c,Pender2012}. Full details of the specific baths treated are available in appendix 3 of the supplementary information, but for the purpose of the following section we will combine them into a single Markovian bath 
% $\mathcal{D}$ 
possessing the two-time correlation function
$C(t,s)=\delta(t-s) D_p$.
We take a high-temperature approximation for this Markovian bath, allowing us to neglect any damping effects it might induce. 
Then this bath adds to eq. \eqref{eq.SpectrIdeal} as a linear term $D_p't$.
The mechanical frequency is given by \cite{Millen2015,Fonseca2016}
\begin{equation}
\omega_m = \frac{V_0\beta Q}{\sqrt{2}m\Omega_dd^2},
\end{equation}
where $Q$  is the number of elementary charges attached to the nanosphere, $\beta$ is a form factor of the trap geometry, and $V_0$ is the amplitude of the AC voltage applied to the electrodes -- the DC component being set to 0. The mass of the nanosphere is given by  $m$, $d$ gives  the distance to the endcap electrodes and $\Omega_d$ is the driving frequency of the trap. The range of $\omega_m$ which can be reliably tested over is constrained by the stability of the trap \cite{Paul1990}, as well as by physical restrictions on various parameters.

{\it Probing the Spectrum of Electric Field Noise -- }
% We progress now to an example application of this protocol, namely: probing non-Markovian bath with a Gaussian spectrum of the form $\tilde{C}(\nu)= \eta e^{-\frac{(\nu_0-\nu)^2}{2\gamma^2}}$ in which $\eta$   gives the strength of the coupling of the noise field to the nanosphere, $\nu_0$ gives the resonant frequency of the collapse noise, and $\gamma$ gives the width of the spectrum. 
The system described above takes generality as one of its main strengths -- it could be used to analyse any noise source to which it could be coupled to, and over a wide frequency range. By way of example, we now demonstrate how such a system could be used for a particular real world application, which could be achieved with current technology. 

The specific structure of the electric field noise which affects the levitated particles in Paul traps remains unknown \cite{brownnutt2015ion}. Here we demonstrate how our protocol could be utilised to reconstruct it. A simpler but less realistic example, where a noise with a purely Gaussian spectrum is considered, is given in appendix 4 of the supplementary information.
% \ref{gaussapp}.
Taking achievable parameters for experimental factors (such as background gas pressure, electric field noise, and environment temperature), and using an example of a $R=50$ nm sphere with the density of silica $\rho=2300$ kg/m$^3$ and a charge of $Q=10^3$ $e$  we can expect a heating rate from conventional sources of $D'_p\leq100$ phonon/s at frequencies higher than $10^{3}$ Hz.

\begin{figure}
    \centering
    \includegraphics[width=.45\textwidth]{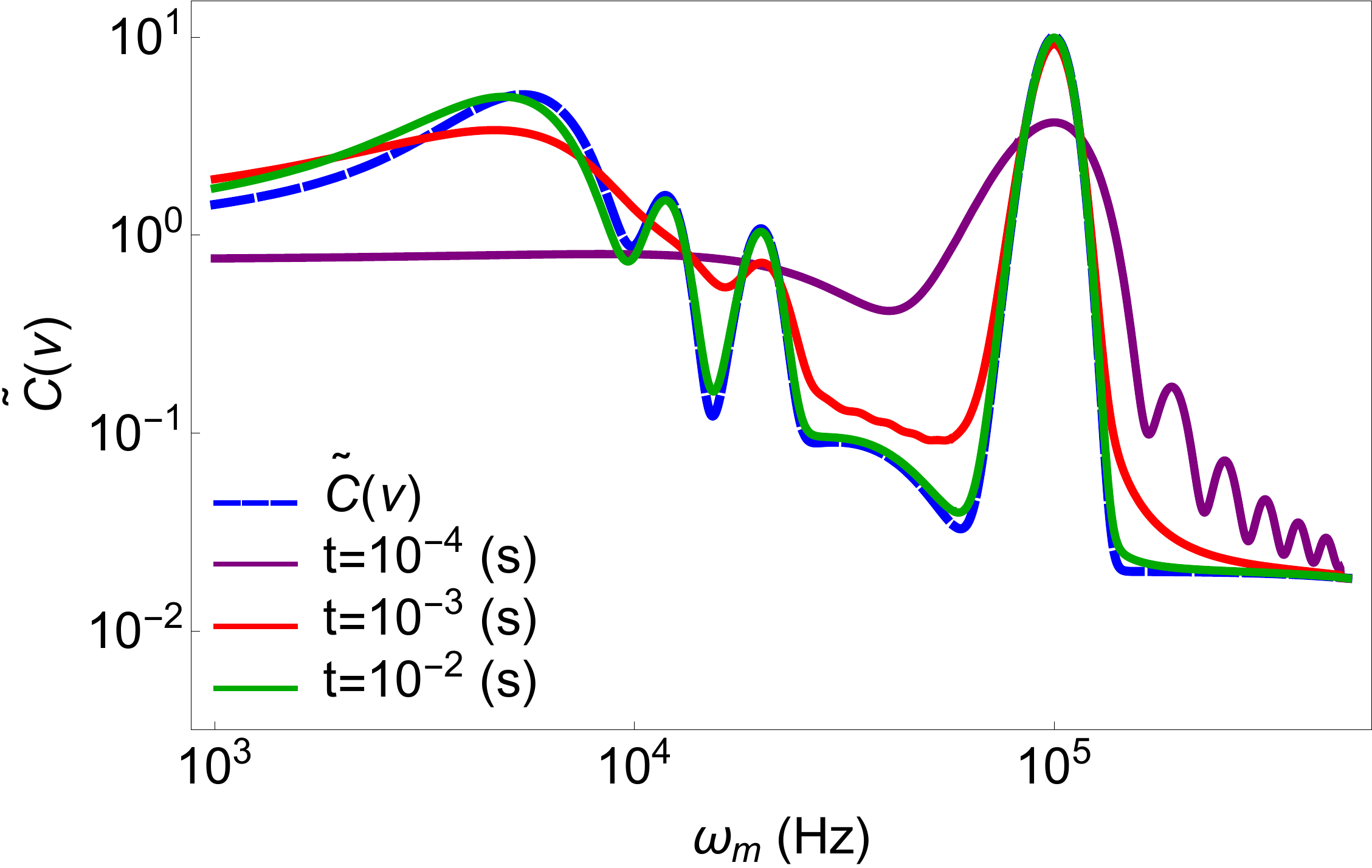}
    \caption{\textbf{Colour Online:} Simulated performance of the spectrometer as enacted on a levitated nanosphere system with realistic parameters, deployed to measure a fictional and non-trivial structure to the electric field noise. The dashed blue line gives the noise spectrum to be measured, $\tilde{C}(\nu)$. An ideal spectrometer would capture this perfectly. The three solid lines show what an experimenter using our protocol would reconstruct, with the purple, red and green lines showing reconstructions using measurement times of $t=10^{-4}$ s, $t=10^{-3}$ s, $t=10^{-2}$ s respectively. The longer measurement times allow for a higher fidelity reconstruction of the objective function -- in particular, the features at lower frequencies. The parameters utilised for the model are achievable with present day technology, and the scaling of the $E$ field noise has been chosen such that even with the noise density structure shown it would yield heating rates commensurate with those typically reported in the literature. 
    %-- which is to say that the Paul trap electric field noise \emph{could} have this structure, and if it did our technique would recover it as seen in the solid line plots. 
    Simulation parameters and the theoretical treatment of the noise sources can be found in  appendix 3 of the supplementary information.}
    %FOR THE 4 K PLOT: T=4K, R=50nm, Q=1000 e,P=10^-9,d=.8mm, 
    \label{fig.NoiseSpectrumWithStructure}
\end{figure}

The basic formulation for the heating rate due to electric field fluctuations in terms of phonons per second is given by \cite{brownnutt2015ion}
% [NEED TO BE DISCUSSED: HOW THE FOLLOWING FORMULA IS DERIVED? IS $S_E(\omega_m)$ THE SAME AS $\tilde{C}(\nu)$ . IF YES I WOULD SKIP THIS eq. AND WRITE DIRECTLY THE NEXT ONE.]
\begin{equation}
    % D_E'=\frac{Q^2}{4 m \hbar \omega_m} S_E(\omega_m),
    D_E'(\omega_m)=\frac{Q^2}{4 m  \omega_m} S_E(\omega_m),
\end{equation}
in which $S_E(\omega_m)$ is the spectral density of the $E$ field noise at the mechanical frequency of the oscillator. In \cite{brownnutt2015ion}, by appraising and comparing a wide range of electrically levitated experiments, Brownnutt et al. attain the following general form for this spectral density:
\begin{equation}
S_E(\omega)=g_E\omega^{-\alpha}d^{-\beta}T^\gamma,    
\end{equation}
in which $g_E$ is a scaling constant, $2 d$ is the inter-electrode distance, and $T$ is the temperature of the electrodes. $\alpha, \beta $ and $\gamma$ are parameters which depend upon trap geometry and experimental specifics. Here, rather than taking $\alpha$ to be a constant, we will replace it with a structure of arbitrary complexity to make $S_E(\omega)  \propto \tilde{C}(\omega)$. 
% g_E d^{-\beta}T^\gamma \tilde{C}(\omega)$.  
Accordingly, our phonon expectation value is given by 
\begin{equation}\label{eq.PhononNumber}
\langle \hat{n} \rangle_t\!=\!\langle \hat{n}\rangle_0+D_p' t
+ \frac{k_E}{2\pi m \omega_m}\int_{-\infty}^{+\infty} \!\!d\nu\, \tilde{C}(\nu)\frac{\sin^2[(\omega_m-\nu)t/2]}{(\omega_m-\nu)^2},
% \langle \hat{n} \rangle_t&=\langle \hat{n}\rangle_0+D_p' t \\
% &+ \frac{Q^2}{2\pi m\hbar \omega_m}d^{-\beta}T^\gamma g_E\int_{-\infty}^\infty \tilde{C}(\nu)\frac{\sin^2[(\omega_m-\nu)t/2]}{(\omega_m-\nu)^2},\nonumber
\end{equation}
in which $D_p'$ gives the effects of all heating sources other than the electric field noise (which, again, are taken as Markovian and without damping), and 
\begin{equation}
    % k_E=\frac{Q^2}{2\pi m\hbar \omega_m}d^{-\beta}T^\gamma g_E
    k_E=Q^2 d^{-\beta}T^\gamma g_E.
\end{equation}
Now, by selecting an appropriate value for $g_E$ we can interrogate how well an arbitrary $\tilde{C}(\nu)$ could be reconstructed in our setting, noting that eq. \eqref{eq.SpecReconst1} would need to be updated to 
\begin{equation}\label{eq.SpecReconstr2}
         \tilde{C}(\omega_m)=\frac{4 m \omega_m}{k_E t}\left(\langle n\rangle_t-\langle n\rangle_0-D_p't\right)
\end{equation}
with the term $D_p't$ accounting for the Markovian bath. Note that $\tilde{C}(\omega_m)$ depends on $\omega_m$ also though $\langle n\rangle_t$, which captures the structure of the spectrum of the noise.

Figure \ref{fig.NoiseSpectrumWithStructure} demonstrates the capabilities of the technique for recovering a hypothetical  complicated structure assigned to the electric field noise (this structure is represented by the dashed blue line). The solid coloured lines show what the experimentalist would reconstruct via the spectrometer, using eq. \eqref{eq.SpecReconstr2}, from the measured values of $\langle n\rangle_t$ as a function of $\omega_m$. The figure gives a clear demonstration of how the resolution changes with the measurement time. The shortest measurement time -- that of the purple line -- recovers only the features at highest frequencies, and with poor fidelity; whereas the long measurement time of the green line gives a faithful reconstruction of the objective function. An additional interesting feature is the ringing effect seen at $\omega_m>10^5$ Hz for the purple line. This ringing effect occurs when there are features in the noise structure sharper than the accuracy $\sim 1/t$ used for reconstructing the spectrum. On the contrary, if the resolution used is high enough, such a ringing is absent. As such, this ringing forms a useful component of the tool. If the experimentalist collects data and infers a particular noise structure from her data, she may wonder if there are finer-grained structures which she might have recovered with longer measurement times. The presence or absence of the characteristic ringing serves to answer her question: if it is there, then the noise contains further features to be discovered. We explain this further in  appendix 5 of the supplementary information. 

The practicability of increasing measurement times inversely proportional to the mechanical frequency makes the technique increasingly viable at higher frequencies; at these high frequencies the accuracy of the reconstruction will be limited more by the sensitivity of the readout and the preparation of the initial state than by the allowed time for evolution. We also note that the electric field noise is particularly distinguishable in this context due to its $Q^2$ scaling, which can be used to disambiguate between the effects of the bath of interest and those of other baths. 

A further example connected to non-Markovian models of spontaneous wave function collapse \cite{Adler2007,Adler2008c,adler2013spontaneous,bassi2014spontaneous,donadi2014spontaneous,Carlesso2018color} is presented in  appendix 2 of the supplementary information, which also contains references \cite{breuer2002theory,Bassi2013b,adler2019testing,fu1997spontaneous,piscicchia2017csl,Smirne2015,nobakht2018unitary,bilardello2016bounds,PhysRevLett.113.020405,Vinante2017,Adamiak2002,Cermak1995a,Asenbaum2013,Romero-Isart2010,Brownnutt2015,Turchette2000,breuer1999stochastic}.

{\it Conclusion --} It is our submission that such the method communicated here would have a broad applicability. Whether for the noise fields invoked by collapse theories, or for more generic non-Markovian quantum noises, this method may be used to detect and characterise the fields of interest. This may be of use in a wide range of technologies, including quantum sensing and improving quantum computing architectures. Further, such a detection and characterisation would shine a clear light for theorists trying to determine the physical origins of such fields -- since the specifics characteristics of a field will carry signatures of its genesis. 

\section{Acknowledgments}
We wish to thanks Prof.~H.~Ulbricht, Prof.~M.~Paternostro, Dr. Brandon Rodenburg, Dr.~A.~Smirne and J. Haase for their valuable suggestions and helpful discussions on this work. D. G. thanks the Centre for Doctoral Training in Controlled Quantum Dynamics. P. B. acknowledges support from the H2020 FET TEQ (grant n. 766900). S.  D.  acknowledges  support  from  Fondazione Angelo Della Riccia, the Foundation BLANCEFLOR Boncompagni Ludovisi, nee Bildt, the Center for Integrated Quantum Science and Technology (IQST), the Frankfurt Institute for Advanced Studies (FIAS) and the Fetzer Franklin Fund.  A. B. acknowledges support from the H2020 FET TEQ (grant n.  766900),  INFN, and the COST action QTSpace.

\clearpage
%%%%%%%%%%%%%%%%%%%%%%%%%%%%%%%%%%%%%%%%%%%%%%%%%%%
%Supplementary

\onecolumngrid
\section*{Supplementary Information}
\setcounter{equation}{0}
\renewcommand{\theequation}{S\arabic{equation}}
\setcounter{section}{0}
\setlength{\parindent}{0pt}
\setlength{\parskip}{1ex}
\count\footins = 1000
\subsection{APPENDIX 1: An Alternative Derivation of the Formalism}
Here we give an alternative derivation for the master equation of the spectrometer, and attendant heating rate. In this version we will opt for ladder operator coupling over position-position coupling, in contrast to the main text. This will allow us to derive all the theory for the spectrometer from the level of the Hamiltonian, without referring to master equations derived elsewhere. This derivation has the advantage of being more intuitive than the one included in the main text and valid also for non-Gaussian states (note, however, that it will not be able to distinguish such non-Gaussian noises from more ordinary Guassian ones). The trade-offs are that the master equation will be derived only perturbatively, and that the interaction Hamiltonian will be equivalent to making the rotating wave approximation (RWA) on the corresponding term in eq.(4) of the main text. 
%Since it attains the same final equations as the derivation in the main text, it is included here for purely pedagogical purposes. 
\paragraph{}

We consider a Hamiltonian which describes a harmonic oscillator coupled to a bosonic bath of independent harmonic oscillators. The free Hamiltonians of the system and the bath are the same as those defined in (2) and (3). However, here we consider an interaction Hamiltonian of the form
\begin{equation}\label{hirwa}
    \hat{H}_I = \hat{a}^\dagger \sum_\alpha g'_\alpha \hat{b}_\alpha +\hat{a}\sum_\alpha g_\alpha^{'*} \hat{b}^\dagger_\alpha
\end{equation}
where $g'_\alpha$ denotes the coupling between the system and a given bath mode. Note that $H_I$ in Eq. (4) reduces to the RWA form of Eq. \eqref{hirwa} when setting $g'_{\alpha}=g_{\alpha}^{'*}=-\frac{ g_{\alpha}}{2\sqrt{m\omega m_{\alpha}\omega_{\alpha}}}=-\frac{1}{2}\sqrt{\frac{m_{\alpha}\omega_{\alpha}^{3}}{m\omega}}$.
\paragraph{}
We now move to the interaction picture as explained in \cite{breuer2002theory}: $\tilde{H}_I(t) = e^{i(\hat{H}_\mathcal{S}+\hat{H}_\mathcal{B})t}\hat{H}_Ie^{-i(\hat{H}_\mathcal{S}+\hat{H}_\mathcal{B})t}$, $\tilde{\rho}(t)=e^{i(\hat{H}_\mathcal{S}+\hat{H}_\mathcal{B})t}\rho(t)e^{-i(\hat{H}_\mathcal{S}+\hat{H}_\mathcal{B})t}$ with an accompanying master equation
\begin{align}
\frac{d}{d t}\tilde{\rho}(t)&=-i[\tilde{H}_I(t),\tilde{\rho}(t)]\label{eq.masterpremarkov}.
% \\
% \shortintertext{which is solved by}
% \tilde{\rho}(t)&=\rho(0)-i\int_0^tds\,[\tilde{H}_I(s),\tilde{\rho}(s)]]
\end{align}
We assume that at $t=0$, the total state factorises as $\tilde{\rho}(0)=\tilde{\rho}_S(0) \rho_E(0)$ where  $\rho_S(t) $ and $\rho_E(t)$ give the reduced state operators for the system and environment respectively. We make the Born approximation, which extends this for all times, and assumes that the environment is left relatively unchanged by its interactions with the system: $\tilde{\rho}(t)\approx\tilde{\rho}_S(t)\rho_E$. This gives an integral solution to eq. (\ref{eq.masterpremarkov}) of the form $\tilde{\rho}(t)=\rho(0)-i\int_0^tds\,[\tilde{H}_I(s),\tilde{\rho}_S(s)\rho_E]$.
%[WHY THE INITIAL CONDITION TERM IS NOT THERE? I THINK BREURER WAS GIVING SOME REASON BASED ON THE BATH. ALSO SHOULD NOT BE $\rho(s)$ INSTEAD OF $\rho(t)$ INSIDE THE INTEGRAL?].
Next, we want to re-insert this solution back into equation (\ref{eq.masterpremarkov}), and then we get
% [MAYBE, WITHPUT GOING IN ALL DETAIL WITH THIS PROCEDURE WHICH IS QUITE STANDARD, SHOULD WE JUST REFER TO BREUER AND MENTION ALL THE ASSUMPTIONS DONE?]
\begin{equation}
\frac{d \tilde{\rho}_S(t)}{d t} =-i{\rm Tr_E}\left[[\tilde{H}_I(t),\rho(0)]\right] -\int_0^tds\,{\rm Tr}_E\left[[\tilde{H}_I(t),[\tilde{H}_I(s),\tilde{\rho}_S(s)\rho_E]]\right]. \label{eq.masterinteraction}
\end{equation}
This process could in principle be repeated ad infinitum, generating integrands containing commutations of ever higher order. The weak coupling assumption however, allows us to assume that such terms will be of diminishing magnitude. Therefore, we follow convention and limit ourselves to the second order terms seen in eq. \eqref{eq.masterinteraction}. By limiting ourselves to the second cumulants of the noise function, we also bar ourselves from being able to reconstruct any non-Gaussian features of the noise -- since these are only present in higher cumulants (for an example of a spectrometer which would be able to reconstruct non-Gaussian features, see \cite{Norris2016,sung2019non}). We can assume the first term on the right hand side, $-i{\rm Tr}_E\left[[\tilde{H}_I(t),\rho_S(0)\rho_E]\right]$, to equal to zero, a condition that is always verified when the environment is in a thermal state.

% \footnotetext[1]{}

We now introduce the grand operators for the bath
\begin{align*}
\hat{B} =  \sum_\alpha g_\alpha' \hat{b}_\alpha,\quad\hat{B}^\dagger = \sum_\alpha g^{*\prime}_\alpha \hat{b}_\alpha^\dagger .
\end{align*}
Note that the $\hat{B}$  operator introduced here is different than the one used in the main text -- here it is related to the annihilation operators, whereas in the main text the bath operators sum over position operators. The grand operators here possess two-time correlation functions
\begin{align}
C_1(t,s)&={\rm Tr}_E[\tilde{B}^\dagger(t)\tilde{B}(s)\rho_E]\label{eq.C1}\\
C_2(t,s)&={\rm Tr}_E[\tilde{B}(t)\tilde{B}^\dagger(s)\rho_E]\label{eq.C2}\\
F(t,s)&={\rm Tr}_E[\tilde{B}(t)\tilde{B}(s)\rho_E]\\
G(t,s)&={\rm Tr}_E[\tilde{B}^\dagger(t)\tilde{B}^\dagger(s)\rho_E].
\end{align}

\paragraph{}
We notice that 
\begin{align*}
C_1(s,t)&={\rm Tr_E}[\tilde{B}^\dagger (s)\tilde{B}(t)\rho_E]={\rm Tr}_E[\left(\tilde{B}^\dagger (t)\tilde{B}(s)\right)^\dagger\rho_E]\\
&={\rm Tr}_E[\tilde{B}^\dagger(t)\tilde{B}(s)\rho_E]^*=C_1^*(t,s)
\end{align*}
and by a similar logic, $C_2(s,t)=C_2^*(t,s)$. We also have 
\begin{align*}
F(s,t)&={\rm Tr}_{E}[\tilde{B}(s)\tilde{B}(t)\rho_{E}]={\rm Tr}_{E}[\left(\tilde{B}(s)\tilde{B}(t)\right)^{\dagger}\rho_{E}]^{*}\\
&={\rm Tr}_{E}[\tilde{B}^{\dagger}(t)\tilde{B}^{\dagger}(s)\rho_{E}]^{*}=G^{*}(t,s)
\end{align*}
and similarly $G(s,t)=F^{*}(t,s)$. From this point on we will take the arguments for these functions as implicit, replacing $C_1(t,s)\to C_1, C_1(s,t)\to C_1^*$ and so forth. 
At this point, a Markovian bath would correspond to taking $C_1 = C_2 = J\delta(t-s)$ and $F = G = 0$. However, our intention is to recover information regarding these functions through the dynamics of the system, and as such we leave them general for now.

Putting the above correlation functions into Eq (\ref{eq.masterinteraction}), expanding and re-arranging we arrive at 
% \begin{widetext}
\begin{align}\label{me223}
    \frac{d \tilde{\rho}_S(t)}{d t}=&-\int_{0}^{t}ds\,\left[F\left[\tilde{a}^{\dagger}(t),\tilde{a}^{\dagger}(s)\tilde{\rho}_{S}(s)\right]+C_{1}\left[\tilde{a}(t),\tilde{a}^{\dagger}(s)\tilde{\rho}_{S}(s)\right]+C_{2}\left[\tilde{a}^{\dagger}(t),\tilde{a}(s)\tilde{\rho}_{S}(s)\right]+G\left[\tilde{a}(t),\tilde{a}(s)\tilde{\rho}_{S}(s)\right]\right.\nonumber\\
    &\left.+G^{*}\left[\tilde{\rho}_{S}(s)\tilde{a}^{\dagger}(s),\tilde{a}^{\dagger}(t)\right]+C_{1}^{*}\left[\tilde{\rho}_{S}(s)\tilde{a}(s),\tilde{a}^{\dagger}(t)\right]+C_{2}^{*}\left[\tilde{\rho}_{S}(s)\tilde{a}^{\dagger}(s),\tilde{a}(t)\right]+F^{*}\left[\tilde{\rho}_{S}(s)\tilde{a}(s),\tilde{a}(t)\right]\right].
\end{align}
% \begin{align}
%     \frac{d\rho(t)}{dt}=&-i[\tilde{H}_I(t),\rho(0)]-\!\!\!\int_0^t \!\!\! ds \Big[ C_{t,s}^1e^{i\omega_m(s-t)}[\hat{a},\hat{a}^\dagger \rho]+C^{1*}_{t,s}e^{-i\omega_m(s-t)}[\rho \hat{a},\hat{a}^\dagger] +C^2_{t,s}e^{-i\omega_m(s-t)}[\hat{a}^\dagger,\hat{a}\rho]-C^{2*}_{t,s}e^{i\omega_m(s-t)}[ \hat{a},\rho \hat{a}^\dagger]\nonumber\\
%     &+F_{t,s}e^{i\omega_m(t+s)}[\hat{a}^\dagger,\hat{a}^\dagger\rho]+F_{t,s}^*e^{-i\omega_m(s+t)}[\rho \hat{a},\hat{a}]+ F_{s,t}e^{i\omega_m(t+s)}[\rho \hat{a}^\dagger,\hat{a}^\dagger]+F_{s,t}^*e^{-i\omega_m(s+t)}[\hat{a},\hat{a}\rho]\Big]
% \end{align}
% \end{widetext}
We now make the first Markov approximation which allows us to make the replacement $\tilde{\rho}_S(s)\to\tilde{\rho}_S(t)$ in eq. (\ref{me223}). This approximation is valid when the dynamics of the bath are much faster than those of the system. The physical reasoning is that if the state of the system is to have a dependence upon its state at a previous time, that dependence must here be mediated by interactions with the environment, which would need to act as a memory imprinted with the prior state of the system.
%the dependence of the system dynamics from its past state is a consequence that at a previous time the system interacted with the environment, modifying its dynamics, and now the modified environment is interacting with the system. 
In other words, it is a back-reaction of the environment from being stimulated by the system at previous times. However, if  the environment is fast enough in resetting its state,  and effectively  loses memory of the interaction with the system quickly, the backreaction cannot occur. Let us be more precise: in eq. (\ref{me223}) the environment enters  the system’s dynamics through the correlation functions $C_1(t,s)$, $C_2(t,s)$, $F(t,s)$, $G(t,s)$. When we talk about a  fast resetting of the dynamics of the bath we precisely mean that the decay time $t_c$ of these functions is much shorter than the typical evolution time of the system. In such a case, the  correlation function $C_1(t,s)$ (and similarly the others) is appreciably different from zero only for values of $s$ such that $t-s\leq t_c$. If the typical time scale of the system dynamics $\tau$  is much longer than $t_c$, one can approximate $\rho(s) C_1(t,s)\simeq \rho(t) C_1(t,s)$ which is  the first Markov approximation. Note that for a thermal bath $t_c=\hbar/(k_B T)$, and for the oscillator $\tau=\omega_m^{-1}$, therefore the condition $\tau\gg t_c$ in this case is equivalent to $\hbar \omega_m\ll   k_B T   $. This is fulfilled when the dissipative effects of the bath are negligible, which is the regime we considered also in the main text, and will again consider below (see paragraph after eq. (\ref{fam})). For a more detailed discussion on the validity of the Markov approximation we refer the reader to \cite{carmichael2010statistical} and the  references therein. As a final remark, we wish to point out that the Markov approximation does \emph{not} equate to Markovian dynamics -- it is a necessary but insufficient condition to attain them\footnotemark{}. 
\footnotetext{Alternatively, one could also yield an expression equivalent to equation \ref{me222} \emph{without} \emph{without} assuming that the timescale of the bath dynamics is much smaller than timescale of the system, by using the time-convolutionless method (more commonly referred to as TCM) described  in ch.9 of  \cite{breuer2002theory} or section 2 of \cite{breuer1999stochastic},  taking the perturbative expansion to second order. We have opted instead for the picture given here, on the grounds that it is more physically intuitive.}

Therefore, using Markov approximation on eq. (\ref{me223}), we get 
\begin{align}\label{me222}
    \frac{d \tilde{\rho}_S(t)}{d t}=&-\int_{0}^{t}ds\,\left[F\left[\tilde{a}^{\dagger}(t),\tilde{a}^{\dagger}(s)\tilde{\rho}_{S}(t)\right]+C_{1}\left[\tilde{a}(t),\tilde{a}^{\dagger}(s)\tilde{\rho}_{S}(t)\right]+C_{2}\left[\tilde{a}^{\dagger}(t),\tilde{a}(s)\tilde{\rho}_{S}(t)\right]+G\left[\tilde{a}(t),\tilde{a}(s)\tilde{\rho}_{S}(t)\right]\right.\nonumber\\
    &\left.+G^{*}\left[\tilde{\rho}_{S}(t)\tilde{a}^{\dagger}(s),\tilde{a}^{\dagger}(t)\right]+C_{1}^{*}\left[\tilde{\rho}_{S}(t)\tilde{a}(s),\tilde{a}^{\dagger}(t)\right]+C_{2}^{*}\left[\tilde{\rho}_{S}(t)\tilde{a}^{\dagger}(s),\tilde{a}(t)\right]+F^{*}\left[\tilde{\rho}_{S}(t)\tilde{a}(s),\tilde{a}(t)\right]\right].
\end{align}

We now want to transition back from the interaction picture to the Schr\"odinger, which we can find using 
\begin{equation}\label{sch_pic}
    \frac{d \rho_S(t)}{d t}=-i\left[\hat{H}_{\mathcal{S}},\rho_{S}(t)\right]+e^{-i\hat{H}_{\mathcal{S}}t}\left(\frac{d}{dt}\tilde{\rho}_{S}(t)\right)e^{i\hat{H}_{\mathcal{S}}t}.
\end{equation}
The exponentials in the last term of eq. (\ref{sch_pic}) acts on the terms in eq. (\ref{me222}) sending $\tilde{\rho}_{S}(t)\longrightarrow\rho_{S}(t)$, $\tilde{O}(t)\longrightarrow \hat{O}(0)=\hat{O}$ and $\tilde{O}(s)\longrightarrow \hat{O}(s-t)$.  Then, using $\tilde{a}(s-t)=\hat{a}\,e^{i\omega_{m}(t-s)}$, one gets:
\begin{align}
    \frac{d \rho_S(t)}{d t}=&-i\left[\hat{H}_{\mathcal{S}},\rho_{S}(t)\right]-\int_{0}^{t}ds\,\left\{Fe^{-i\omega_{m}(t-s)}\left[\hat{a}^{\dagger},\hat{a}^{\dagger}\rho_{S}(t)\right]+C_{1}e^{-i\omega_{m}(t-s)}\left[\hat{a},\hat{a}^{\dagger}\rho_{S}(t)\right]\right.\nonumber\\
    &+C_{2}e^{i\omega_{m}(t-s)}\left[\hat{a}^{\dagger},\hat{a}\rho_{S}(t)\right]+Ge^{i\omega_{m}(t-s)}\left[\hat{a},\hat{a}\rho_{S}(t)\right]+G^{*}e^{-i\omega_{m}(t-s)}\left[\rho_{S}(t)\hat{a}^{\dagger},\hat{a}^{\dagger}\right]\nonumber\\
    &\left.+C_{1}^{*}e^{i\omega_{m}(t-s)}\left[\rho_{S}(t)\hat{a},\hat{a}^{\dagger}\right]+C_{2}^{*}e^{-i\omega_{m}(t-s)}\left[\rho_{S}(t)\hat{a}^{\dagger},\hat{a}\right]+F^{*}e^{i\omega_{m}(t-s)}\left[\rho_{S}(t)\hat{a},\hat{a}\right]\right\}
\end{align}
Now, if we name right hand side term of the above equation to be a Liouvillian super-operator $\frac{d}{dt}\rho_S(t)=\mathcal{L}\rho_S(t)$, then we can examine the time evolution of the average of any operator $\hat{\mathcal{O}}$ of the system via
\begin{equation}
    \frac{d}{dt}\langle\hat{\mathcal{O}}\rangle_t={\rm Tr}[\hat{\mathcal{O}}\mathcal{L}\rho_S(t)]. 
\end{equation}
Using this, after some algebra we can find an equation of motion for the number operator $\hat{n}=\hat{a}^\dagger \hat{a}$:
% \begin{widetext}
\begin{align}
\frac{d}{dt}\langle\hat{n}\rangle_{t}=\int_{0}^{t}ds\,&\left[\left(C_{1}e^{-i\omega_{m}(t-s)}+C_{1}^{*}e^{i\omega_{m}(t-s)}\right)\langle\hat{a}\hat{a}^{\dagger}\rangle_{t}-\left(C_{2}e^{i\omega_{m}(t-s)}+C_{2}^{*}e^{-i\omega_{m}(t-s)}\right)\langle\hat{a}^{\dagger}\hat{a}\rangle_{t}\right.\nonumber\\
&\left.-\left(F-G^{*}\right)e^{-i\omega_{m}(t-s)}\langle\hat{a}^{\dagger}\hat{a}^{\dagger}\rangle_{t}-\left(F^{*}-G\right)e^{i\omega_{m}(t-s)}\langle\hat{a}\hat{a}\rangle_{t}\right]\label{eq.EqmNumberOp1}
    % \frac{d}{dt}\langle \hat{a}^\dagger \hat{a}\rangle =\int_0^s ds\, &\left(\langle \hat{a}^\dagger \hat{a} \rangle+1\right) \Big( C^1_{t,s} e^{i \omega_m(s-t)}+C^{1*}_{t,s}e^{-i\omega_m(s-t)}\Big)
    % -\langle \hat{a}^\dagger \hat{a}\rangle\Big( C^2_{t,s}e^{-i\omega_m(s-t)}+C^{2*}_{t,s}e^{i\omega_m(s-t)}\Big)\\
    % &+F_{t,s}e^{i\omega_m(s+t)}\langle \hat{a}^\dagger \hat{a}^\dagger \rangle+F_{t,s}^*e^{-i\omega_m(s+t)}\langle \hat{a}\hat{a}\rangle -F_{s,t}e^{i\omega_m(s+t)}\langle \hat{a}^\dagger \hat{a}^\dagger\rangle-F_{s,t}^*e^{-i\omega_m(s+t)}\langle \hat{a} \hat{a}\rangle\nonumber
\end{align}
% \end{widetext}
% \shortintertext{Now, if $C_{1,2} \in \mathbb{R}$ then we have}
%     \frac{d}{dt}\langle n(t)\rangle&=2(\langle n(t)\rangle +1) D_1-2\langle n(t)\rangle\\
%     D_2\nonumber
If we make the assumption that the bath correlations are symmetric with respect time reversal and translation  i.e. that $X(t,s)=X(|t-s|)$ with $X=C_{1},C_{2},F,G$, one immediately finds  $C_i=C_i^*$ for $i=1,2$ and $F=G^{*}$. Then, eq. (\ref{eq.EqmNumberOp1}) simplifies to
\begin{align}
    \frac{d}{dt}\langle n \rangle_t = 2\int_{0}^{t}ds\,\cos\left[\omega_{m}(t-s)\right]\left[C_{1}+\left(C_{1}-C_{2}\right)\langle\hat{n}\rangle_{t}\right]. \label{eq.EqmNumberOp2}
\end{align}

Proceeding similarly to what we have done in the main text, we use 
\begin{equation}
    \int_0^t ds\,C_i(t-s)\cos[\omega_m(t-s)]=\frac{1}{2}\int_{-t}^t dy\,C_i(y)e^{i\omega_m y}.
\end{equation}
% Now, taking the example of the $C_1$ term:
% \begin{alignat*}{2}
%     &\int_0^tds\, C(s-t)\cos\omega(s-t)\\
%     &=\frac{1}{2}\int_0^tdyC(y)(e^{i\omega y})+e^{-i\omega y}\\
%     \shortintertext{where $y=s-t$}
%     &=\frac{1}{2}\int_0^tC(t)e^{i\omega y}+\frac{1}{2}\int_0^tC(y)e^{-i\omega y}\\
%     \shortintertext{Now for the second term introduce $y'=-y$}
%     & -\frac{1}{2}\int_0^{-t}dy'\,C(-y)e^{i\omega y'}\\
%     \shortintertext{Now recalling that $C(-x)=C(x)$ we have}
%     &-\frac{1}{2}\int_{-t}^0dy'\,C(y')e^{i\omega y}\\
%     &=\frac{1}{2}\int_{-t}^tC(y)e^{i\omega y}
% \end{alignat*}

Next, we Fourier expand $C(y)$ as follows:
\begin{align}
C_{i}(y)=\frac{1}{2\pi}\int_{-\infty}^{+\infty}d\nu\,\tilde{C}_{i}(\nu)e^{-i\nu y} 
\end{align}
with which we re-write Eq. \eqref{eq.EqmNumberOp2} as
\begin{equation}
\frac{d}{dt}\langle\hat{n}\rangle_{t}=\frac{1}{2\pi}\int_{-\infty}^{+\infty}d\nu\,\int_{-t}^{t}dy\,\left[\tilde{C}_{1}(\nu)+\left(\tilde{C}_{1}(\nu)-\tilde{C}_{2}(\nu)\right)\langle\hat{n}\rangle_{t}\right]e^{i(\omega_{m}-\nu)y}
\end{equation}

Introducing here $\tilde{C}_3(\nu)=\tilde{C}_2(\nu)-\tilde{C}_1(\nu)$ and integrating over $y$ we get
\begin{align}
\frac{d}{dt}\langle\hat{n}\rangle_{t}=\frac{1}{\pi}\int_{-\infty}^{+\infty}d\nu\,\frac{\sin\left[(\omega_{m}-\nu)t\right]}{(\omega_{m}-\nu)}\left[\tilde{C}_{1}(\nu)-\tilde{C}_{3}(\nu)\langle\hat{n}\rangle_{t}\right]\label{eq.altDevPenult}
\end{align}

which fits the familiar form
\begin{equation}\label{fam}
    \frac{d}{dt}\langle \hat{n}\rangle_t =A(t)-\gamma \langle n\rangle_t.
\end{equation}
% in which $A(t) = \int_{-\infty}^\infty d\nu\, \sin \[t (\omega-\nu)\]/(\omega-\nu)C_1(\nu),\quad \gamma=$
Here, $\gamma$ provides a damping effect. At this point we make a further assumption: we  neglect the damping effect by setting $\gamma=0$, entailing that  $\tilde{C}_1(\nu)=\tilde{C}_2(\nu)=\tilde{C}(\nu)$. If the bath were strictly thermal, this would be equivalent to taking it to be of infinite temperature. However here we have still not set the specific form of the spectrum of the bath, which remains general (barring the assumptions made). Therefore, neglecting $\gamma$ should be read as meaning that in the regime we are considering the harmonic oscillator energy is much lower than the energy the system will have when in equilibrium with the environment, and that as such the damping rate is negligible compared to the heating rate.

Making this assumption, eq. \eqref{eq.altDevPenult} then reduces to eq. (12) in the main text, barring a factor of $4\pi m\omega_m$, which is owed to the different definitions of the grand bath operators used here and in the main text (here we sums over ladder operators, there over position operators). This derivation then leads as a final result exactly to eq. (13) from the main text.

\subsection{APPENDIX 2: The Non-White Noise Continuous Spontaneous Localization  (CSL) Model}
Another particularly interesting application of this protocol would be the first experimental investigation of general non-white models of spontaneous wavefunction collapse. For the reader who is unfamiliar with them, a thorough overview can be found in \cite{Bassi2013b} -- however, for the purpose of this work, it will also suffice to think of them as simply introducing a noise source to the dynamics of our harmonic oscillator. In their original formulation, the noises in collapse models have white spectra and, under this assumption, a wide range of experiments have been  considered over the last two decades for testing these models. However, when the noise is modelled in a more realistic way having some non-white spectrum -- with, for example, a cutoff at high frequencies -- the predictions from these models change \cite{adler2019testing,carlesso2018colored} and some of these experimental tests e.g. the study of X-rays emission from matter \cite{fu1997spontaneous,adler2013spontaneous,bassi2014spontaneous,donadi2014spontaneous,piscicchia2017csl}, lose their effectiveness. Therefore, an experimental setup which can efficiently test noises with generic spectrum is significant for testing non-white noise collapse models. The noise proposed by such models fits the criteria for being properly investigable through the methods outlined here -- it weakly interacts with the system, so we are in the weak coupling regime; and it is taken to have an infinite temperature\footnotemark[1]{}, which accords with our decision to neglect the damping effects. 
%it is subject to the Markov approximation, entailing that the dynamics of its `bath' evolve on a timescale much faster than that of the sytem under study
\footnotetext[1]{The model put forward in \cite{Smirne2015} describes a variant theory with a \emph{finite} temperature noise. We expect that for temperature of the noise comparable with the one of the system the method outlined here will require major modifications. For a discussion on experimental tests of the dissipative version of CSL model using cold atoms and optomechanical systems see \cite{nobakht2018unitary,bilardello2016bounds}.  However, as long as the temperature of the noise is much larger than the one of the system, dissipative effects can be neglected and the methods outlined here may be used.}

The CSL model master equation, when non-white noises are considered, takes the form \cite{Adler2007,Adler2008c}:
\begin{equation}\label{cslme}
\frac{d\rho\left(t\right)}{dt}=-i\left[\hat{H},\rho\left(t\right)\right]-8 \lambda_{\rm csl} \pi^{\frac{3}{2}}r_c^3\!\int\!\! d\mathbf{z}\!\int_{0}^{t}\!\!ds\,C\!\left(t,s\right)\left[\hat{M}_{\mathbf{z}},\left[\hat{M}_{\mathbf{z}}\left(s-t\right)\!,\!\rho\left(t\right)\right]\right]
\end{equation}
where, for a system composed of $N$ particles in the  first quantization picture, 
\begin{equation}\label{Mz}
\hat{M}_{\mathbf{z}}=\frac{1}{m_{0}}\sum_{j=1}^{N}m_{j}g\left(\mathbf{\hat{x}}_{j}-\mathbf{z}\right)
\end{equation}
and 
\begin{equation}\label{g}
g\left(\mathbf{x}\right)=\frac{1}{\left(\sqrt{2\pi}r_{c}\right)^{3}}e^{-\frac{\mathbf{x}^{2}}{2r_{c}^{2}}},
\end{equation}
with $m_j$ the mass of the $j$-th particle, $m_0$ a reference mass chosen equal to the mass of the nucleon. $C(t,s) = \langle \xi(t)\xi(s)\rangle$ is the autocorrelation function of the noise field $\xi$ which is introduced in CSL for inducing the collapse of the wave function. The parameter $\lambda_{\rm csl}$ in front of the second term of Eq. \eqref{cslme} sets the strength of the interaction with the collapse noise and has dimension of Hz, while the parameter $r_C$ sets the typical correlation length of the noise field, i.e. noises associated to points in space more distant than $r_C$ are uncorrelated. Together, these two parameters characterise the standard white noise CSL model, in which $C(t,s) = \delta(t-s)$.
% In the next sections we will also consider the parameter $\lambda=\gamma/(8\pi^{3/2}r_{C}^{3})$, which has the dimensions of a rate and, together with $r_C$, it is the parameter on which most of the literature refers to set the bounds on standard CSL model. 
We also recall that the master equation \eqref{cslme} is an approximate equation where only first order contribution of $\lambda_{\rm csl}$ is considered, as discussed in detail in \cite{Adler2007,Adler2008c}. However, since the parameter $\lambda_{\rm csl}$ is typically small (in fact, as mentioned above, we are in the weak coupling regime), this first order perturbation expansion is a good approximation of the exact master equation.
In the case of a rigid, spherical object (such as the nanosphere we are considering), the master equation \eqref{cslme} can be simplified so as to fit into the case which interests us; that which describes the motion of a levitated nanosphere. 

First of all it is convenient to introduce the Fourier transform of the Gaussian in Eq. \eqref{g}. Taking the Fourier transform allows us to rewrite Eq. \eqref{Mz} as follows: 
\begin{equation}\label{Mzfourier}
\hat{M}_{\mathbf{z}}=\frac{1}{(2\pi)^{3}m_{0}}\sum_{j=1}^{N}m_{j}\int d\mathbf{k}e^{-\frac{1}{2}\mathbf{k}^{2}r_{c}^{2}}e^{i\mathbf{k}\cdot(\hat{\mathbf{x}}_{j}-\mathbf{z})}.
\end{equation}
Substituting this into master equation \eqref{cslme}, we obtain
\begin{align}\label{cslmefourier}
\frac{d\rho\left(t\right)}{dt}&=-i\left[\hat{H},\rho\left(t\right)\right]-\frac{\lambda_{\rm csl} r_c^3}{\pi^{3/2}m_{0}^2}\sum_{i,j=1}^{N}m_{i}m_{j}\int d\mathbf{k}e^{-\mathbf{k}{}^{2}r_{c}^{2}}\nonumber\\
&\times\int_{0}^{t}ds\,C\left(t,s\right)\left[e^{-i\mathbf{k}\cdot\hat{\mathbf{x}}_{i}},\left[e^{i\mathbf{k}\cdot\hat{\mathbf{x}}_{j}\left(s-t\right)},\rho\left(t\right)\right]\right].
\end{align}
We now implement the rigid body approximation. In such a case, the position of each particle can be written as $\hat{\mathbf{x}}_{j}\simeq\hat{\mathbf{x}}+\mathbf{x}_{j}^{(0)}$, where $\mathbf{x}_{j}^{(0)}$ is a fixed position around which the $j$-th particle oscillates while $\hat{\mathbf{x}}$ describes the center of mass oscillations. It is also convenient to introduce the mass density
\begin{equation}\label{massden}
\mu(\mathbf{y})=\sum_{i=1}^{N}m_{i}\delta(\mathbf{y}-\mathbf{x}_{i}^{(0)})
\end{equation}
so that the following relation holds: 
\begin{equation}
\sum_{i=1}^{N}m_{i}e^{-i\mathbf{k}\cdot\mathbf{x}_{i}^{(0)}}=\int d\mathbf{y}\mu(\mathbf{y})e^{-i\mathbf{k}\cdot\mathbf{y}}=\tilde{\mu}(\mathbf{k})
\end{equation}
and the master equation \eqref{cslmefourier} can be rewritten as

\begin{align}\label{cslmefourier2}
\frac{d\rho\left(t\right)}{dt}&=-i\left[\hat{H},\rho\left(t\right)\right]-\frac{\lambda_{\rm csl} r_c^3 }{\pi^{3/2}m_{0}^{2}}\int d\mathbf{k}e^{-\mathbf{k}{}^{2}r_{c}^{2}}|\tilde{\mu}(\mathbf{k})|^{2}\nonumber\\
&\times\int_{0}^{t}ds\,C\left(t,s\right)\left[e^{-i\mathbf{k}\cdot\hat{\mathbf{x}}},\left[e^{i\mathbf{k}\cdot\hat{\mathbf{x}}\left(s-t\right)},\rho\left(t\right)\right]\right].
\end{align}
% where we have used $\tilde{\mu}(-\mathbf{k})=\tilde{\mu}^{*}(\mathbf{k})$ which follows from the fact that the mass density $\mu(\mathbf{y})$ is a real quantity.
Finally, if we chose our reference system with the origin in the classical center of mass position of the system and we focus on the case where the oscillations of the nanosphere are much smaller than $r_C$ i.e. when:
\begin{equation}\label{cond}
\langle\hat{\mathbf{x}}\left(t\right)^2\rangle\ll r_{C}^2
\end{equation}
we can expand the exponential inside the double commutator of Eq. \eqref{cslmefourier2} and we get
\begin{align}\label{cslmeexpand}
\frac{d\rho\left(t\right)}{dt}&=-i\left[\hat{H},\rho\left(t\right)\right]-\frac{\lambda_{\rm csl} r_c^3 }{\pi^{3/2}m_{0}^{2}}\int d\mathbf{k}e^{-\mathbf{k}{}^{2}r_{c}^{2}}|\tilde{\mu}(\mathbf{k})|^{2}\nonumber\\
&\times\int_{0}^{t}ds\,C\left(t,s\right)\left[\mathbf{k}\cdot\hat{\mathbf{x}},\left[\mathbf{k}\cdot\hat{\mathbf{x}}\left(s-t\right),\rho\left(t\right)\right]\right].
\end{align}
Usually, in experiments we are focused on the motion in one direction, for example the direction along the $x$-axis (which we will denote as ``$q$" for consistency with the notation used in the main text). In such a case eq. \eqref{cslmeexpand} becomes:
\begin{equation}\label{cslmefinal}
\frac{d\rho\left(t\right)}{dt}=-i\left[H,\rho\left(t\right)\right]-\eta_{z}\int_{0}^{t}ds\,C\left(t,s\right)\left[\hat{q},\left[\hat{q}\left(s-t\right),\rho\left(t\right)\right]\right]
\end{equation}
where 
\begin{equation}\label{etaz}
\eta_{z}:=\frac{\lambda_{\rm csl} r_c^3 }{\pi^{3/2}m_{0}^{2}}\int d\mathbf{k}e^{-\mathbf{k}{}^{2}r_{c}^{2}}|\tilde{\mu}(\mathbf{k})|^{2}k_{z}^{2}.
\end{equation}
For the case of a sphere, $\eta_{z}$ becomes
\begin{equation}\label{etazsph}
\eta_{z}=\frac{M^{2}}{m_{0}^{2}}3\lambda\frac{r_{C}^{2}}{R^{6}}\left[R^{2}-2r_{c}^{2}+e^{-\frac{R^{2}}{r_{c}^{2}}}\left(R^{2}+2r_{c}^{2}\right)\right],
\end{equation}
which is consistent with the form factor $\alpha$ computed in \cite{Nimmrichter2014} (defined in Eq. (1) of the paper and computed in Eq. (S10) of its supplementary material) keeping into account that $\eta_{z}=\frac{\lambda}{r_{C}^{2}}\alpha$.

In the case of an harmonic oscillator with natural frequency $\omega$ we have:
\begin{equation}
\hat{q}\left(t\right)=\cos\left(\omega t\right)\hat{q}+\frac{\sin\left(\omega t\right)}{m\omega}\hat{p}
\end{equation}
and the above equation becomes
\begin{equation}\label{eq.cslmefinal3}
\frac{d\rho\left(t\right)}{dt}=-i\left[\hat{H},\rho\left(t\right)\right]-\eta_{z}A(t)\left[\hat{q},\left[\hat{q},\rho\left(t\right)\right]\right]-\eta_{z}B(t)\left[\hat{q},\left[\hat{p},\rho\left(t\right)\right]\right]
\end{equation}
where 
\begin{align}\label{eq.AB}
&A(t)=\int_{0}^{t}ds\,C\left(t,s\right)\cos\left[\omega(s-t)\right],\\
&B(t)=\frac{1}{m\omega}\int_{0}^{t}ds\,C\left(t,s\right)\sin\left[\omega(s-t)\right].
\end{align}

% \begin{equation}\label{eq.cslmefinal2}
% \frac{d\rho\left(t\right)}{dt}=-i\left[\hat{H},\rho\left(t\right)\right]-\eta_{z}A(t)\left[\hat{q},\left[\hat{q},\rho\left(t\right)\right]\right]\nonumber-\eta_{z}B(t)\left[\hat{q},\left[\hat{p},\rho\left(t\right)\right]\right]
% \end{equation}
% with
% \begin{equation}\label{etasphere2}
% \eta_{z}=\frac{M^{2}}{m_{0}^{2}}3\lambda\frac{r_{C}^{2}}{R^{6}}\left[R^{2}-2r_{c}^{2}+e^{-\frac{R^{2}}{r_{c}^{2}}}\left(R^{2}+2r_{c}^{2}\right)\right]
% \end{equation}
% giving the effective coupling between the system and the noise field, and where
% \begin{align}\label{eq.AB2}
% A(t)&=\int_{0}^{t}ds\,C\left(t,s\right)\cos\left[\omega_m(s-t)\right],\\
% B(t)&=\frac{1}{m\omega_m}\int_{0}^{t}ds\,C\left(t,s\right)\sin\left[\omega_m(s-t)\right].
% \end{align}

Eq. \eqref{eq.cslmefinal3} has exactly the same structure as Eq. (8) in the main text, meaning that the analysis would proceed  exactly in the same way as laid out there, and the formula for $\langle n\rangle_{t}$ will be: 
\begin{align}\label{ncsl}
\langle \hat{n}\rangle_{t}=\langle \hat{n}\rangle_{0}+\frac{\eta_{z}}{2\pi m\omega_{m}}\int_{-\infty}^{\infty}d\nu\,\tilde{C}(\nu)\frac{\sin^{2}[(\omega_{m}-\nu)t/2]}{(\omega_{m}-\nu)^{2}}.
\end{align}  
%This means that the scene depicted in figure \ref{plot.CorrelationCompare} could be interpreted in a more specific way. A CSL model using parameters $r_c=10^-7$ m, $\lambda_{\rm csl}=10^{-22}$ Hz\footnotemark[1]{}, and possessing a Gaussian correlation function proscribed by $\gamma=10^4$ Hz and focused at $\nu_0=10^5$ Hz would produce precisely the plot seen in the figure. 
As such, this scheme can be said to offer a concrete way to probe collapse models with arbitrary spectra within the range of the oscillator. This is useful on two counts. First, that it might be used to test \emph{specific} non-white models, which predict particular spectra. And secondly, that if collapse effects are indeed detected \cite{Vinante2017}, the technique communicated here could be used to determine what the \emph{otherwise arbitrary} spectrum of the detected noise field looks like. This in turn would help guide the development of a theory in agreement with the spectrum -- the identification of some physical process capable of giving rise to such a phenomenon. Such data would massively restrict -- and thus enhance --  the search for a theory of wavefunction collapse replete with an ontology for the collapsing field. 
% \footnotetext[1]{This might sound strangely low for a value of $\lambda_{\rm csl}$. However, the comparison to the white noise models with which the reader may be more familiar is not quite so simple, since here the coupling is $\propto \lambda_{\rm csl}/\gamma$, rather than $\propto\lambda_{\rm csl}$}

 %
\subsection{APPENDIX 3: Noise sources and Limitations}
In the main text we mention we mention that the mechanical frequency range for the levitated nanosphere is constrained by physical considerations. Specifically, $V_0$ is limited by the capabilities of the voltage source; $d$ by both considerations of noise and trapping practicalities; and $Q$ is limited both by the fact that only a certain amount of charge can fit on the surface of a nanosphere before the Coulomb potential exceeds the binding energy \cite{Adamiak2002}, and by the methods of affixing that charge \cite{Cermak1995}. For a sphere of radius $R=1\,\mu$m, charges in excess of $Q=10^6$ $e$ have been achieved \cite{Asenbaum2013}.

The example experimental scenario used in the main paper is that described in \cite{Goldwater2018}; a charged silica nanosphere, of radius $R=50$ nm, levitated in an electric Paul trap at a low pressure $P= 10^{-9}$ Pa, and cooled to low initial quanta ($n_0=10$) using the all-electrical cooling methods outlined in \cite{Goldwater2018}.  We take the environment to be at $T=4$ K, and the charge of the sphere to be $Q=1000$ \emph{e}, and the mean distance from the particle to the electrodes to be 0.8 mm. By using the Paul trap and all electrical cooling, we are able to avoid the optical scattering noise which is the scourge of levitated optomechanical experiments \cite{Chang2010}. The remaining noise sources are described below. 
\subsubsection*{Gas Collisions}
The diffusion coefficient due to the collisions with the particles of the gas is \cite{rodenburg2016quantum,Romero-Isart2010a}:
\begin{equation}\label{Dgapp}
D_{g}=\frac{6m_{g}\bar{v}P\pi R^{2}}{\hbar^{2}}=\frac{6\pi PR^{2}}{\hbar^{2}}\sqrt{3m_{g}k_{B}T}    
\end{equation}
where we used the equipartition theorem $\bar{v}=\sqrt{3k_{B}T/m_{g}}$. Eq (\ref{Dgapp}) follows from Eq. (F.9) of \cite{Romero-Isart2010a} taking into account that the parameter $\Lambda$ introduced there is related to the diffusion coefficient via $D_{g}=2\Lambda$. Using eq (\ref{Dgapp}) and the relation $D_{g}'=\frac{\hbar}{2m\omega_{m}}D_{g}$ (compared to the main text, we reintroduced the $\hbar$ in the numerator) one gets $D_{g}'=\frac{6.32\times10^{4}}{\omega_{m}}$, which guarantees that for frequencies higher than $10^{3}$ Hz one has $D_{g}'\leq 100$.

It is worth noting that at the pressures we are considering, the average time between collisions with the background gas will often exceed the measurement period. Nonetheless, the net effects of these collisions must be considered, as they will impact the overall statistics of repeated experimental runs. 
\subsubsection*{Blackbody Radiation}
The contribution from blackbody radiation is comprised of two components; emissive and absorptive noise. The emissive rate from a spherical object in the Rayleigh regime is calculated by Chang et. al in  \cite{Chang2010} to be $D'_{\rm bb} = \frac{2\pi^4}{63}\frac{(k_BT)^6}{c^5\hbar^5\rho \omega}{\rm Im}\frac{\epsilon-1}{\epsilon+2}$; where $\epsilon$ is the permittivity of the object and $\rho$ the density \cite{Chang2010}. Taking into account that for silica $\rho=2330$ Kg/m$^3$, ${\rm Im}\frac{\epsilon-1}{\epsilon+2}\simeq 0.1$ and we are considering $T=4$ K, this value is approximately $10^{-14}\;\omega^{-1}$, which is negligible compared to the noise from gas collisions or the electric field. 

% When considering absorptive noise however, our case is quite different to that of Chang et. al in, since we must consider the blackbody radiation from the nearby surface of the trapping electrode, and not just free space. An analysis of environmental blackbody noise above trap electrodes in \cite{brownnutt2015ion}, analysing both surface emissions and those from free space, yields a hueristic heating rate of 
% \begin{equation}
% D_{\rm bba} = \frac{Q^2 T}{6  d^2 }\times 10^{-27},
% \end{equation}
% which gives an effect which is also negligible compared to our other noise sources. 

\subsubsection{Electric Field Noise}

The models built up around electric field noise in quadrupole traps are a mixture of heuristic description and theory. In general, such models are tested by measuring the heating rate of an ion held in the trap operating in the MHz range. One of the difficulties in attempting to build a coherent model of such noise is that when measuring the noise via the heating rate of an ion, it is impossible to distinguish between the different origins of the noise. As outlined in \cite{brownnutt2015ion}, there are many possible sources of noise, including patch potentials on the surface of the electrodes, Johnson-Nyquist noise, and interference with the equipment from other fields in the lab. 

When modelling our electric field noise we face a dual difficulty. First, the noise sources - as far as they are understood - have primarily been studied in the MHz range, whereas we are interested in a wider range. Secondly, the variation in experimentally detected values for heating effects from electric field noise and the lack of a complete and coherent theoretical framework for treating such noise makes a complete model impossible; either a heuristic one based on data from other experiments or a predictive one based on theory. Instead of attempting such a complete model, we instead work with the generic form \cite{brownnutt2015ion}
\begin{equation}
S_E(\omega) =  g_E \omega^{-\alpha}d^{-\beta}T^\chi.
\label{eq.Enoisespec}
\end{equation}
%
% Following the convention in the literature, we model the electric field noise as white, an approximation which is legitimate when the correlation time of the field is considerably shorter than the heating time for the oscillator (as will be the case in our considerations.)[WILL BE OR IT IS? IF YES HOW THIS IS SUPPORTED BY DATA?]
%
In which $\alpha,\,\beta$ and $\chi$ are parameters to be fit to the specific trap, and $g_E$ is a scaling factor. A trap with such an elecric field noise spectrum will heat the particle at a rate \cite{Turchette2000}
\begin{align}
    D_E(\omega_m)&=\frac{\pi}{2}Q^2S_E(\omega_m)\\
    \shortintertext{in terms of Watts, or}
    D_E'(\omega_m)&=\frac{Q^2}{4m\hbar\omega_m}S_E(\omega_m)\,\,
\end{align}
in terms of phonons per second. Now, in the main text we needed to replace the generic Ohmic structure of $S_E(\omega)$ with something more structured, in order to model the capabilities of the spectrometer. In order to do this, we must first select generic parameters for the trap to ensure that we are modelling a realistic situation. We start from the Ohmic case with $\alpha = 1$ and we set $\beta = 3, \,\chi = 0.57$, and fix $g_E =1.55\times10^{-17}$, which corresponds to a heating rate of 1 quanta per second for a Ca$^{40}$ ion trapped at $\omega = 2 \pi \times 5.5$ kHz -- this models our trap as fairly typical in terms of its electric field noise. From this point, we are able to experiment with different structures for $S_E(\omega)$, provided that it is scaled in such a way as to keep the net heating realistic.

\subsubsection*{Rotational Dynamics}
Another effect of the rf field is worth taking note of however. The heating discussed above is due to noise on the rf field acting on the center of mass motion of the particle, but there is also another mechanism capable of transferring heat to the mechanical frequency of the particle; In the case of a trapped nanosphere the charge can be distributed anisotropically over the surface of the sphere, unlike a single ion. An anisotropic charge distribution can lead to a torque on the particle as it passes through the rf field gradient, which can induce rotation. The energy from this rotation can then couple into the mechanical frequency, causing heating. This effect was explored in \cite{Goldwater2016}, in which we found it to be negligible via numerical simulation. 

\subsubsection{Distinguishing the sources of heating} \label{sec.distinguishing}
In \cite{Goldwater2016}, it was explored how parametric variations could help distinguish between genuine collapse effects and those of ordinary environmental decoherence. Here we will use similar methods, supplemented with the ability to vary the mechanical frequency described above. 

The electric field noise which dominates at higher frequencies has a number of parameters to be characterised; $\alpha,\beta,\chi$ and $g_E$. It may also be the case that the model given in Eq. (\ref{eq.Enoisespec}) is insufficient, and that we need to consider a number of electric field noise spectra which sum together, each with their own values for the above parameters. Nonetheless, electric field noise of any stripe will in principle be easily distinguishable from other noises because of its $Q^2$ scaling. The noise arising from gas collisions will similarly be characterisable by its linear scaling with pressure. 

The net heating rate will be measured through repeated experiments. The values of $D_p$ and $D_E$ can then be specified through the parametric variations which will isolate them. Once these are determined, and remaining un-explained heating would become a candidate, for example, as evidence of collapse models noise. The spectrum of this would then be probed, as outlined above. The main method for distinguishing it as genuine collapse noise would then be to vary the radius $R$ of the sphere, which is predicted to elicit something akin to an effective resonance from the noise field when $R\approx r_c$. This effect was first shown in \cite{Nimmrichter2014}, and our results in  \cite{Goldwater2016} concur. 

% For silica silica sphere this would be approximately $\phi=5\times 10^8$ V/m, which would give a maximum charge of 

% \begin{equation}
% Q_{\rm n max} = \frac{\pi \epsilon_0 R^2 E_l}{e},
% \end{equation}
% in which  $\epsilon_0$ is the vacuum permittivity and $R_s$ is the radius of the sphere. 

% The lower limit on the mechanical frequency is approximately 10 Hz, whereas the maximum  frequency is determined by (TO BE FILLED IN). 

% As can be seen in figure \ref{plot.heatingrates}; beyond frequencies of $\sim 10^7$ Hz, the gas noise is high enough to preclude useful investigation, whereas the electric field dominates at lower frequencies. As demonstrated in the Supplementary Material (SM), the dominant sources of heating in this specific scenario are collisions with the background gas and noise on the trapping electric field.  
% \begin{figure}
% 	\centering
% 	\includegraphics[width=.45\textwidth]{./Plots/HeatingRates2.pdf}
% 	\caption{Heating rates (given in phonons / second) from environmental sources. $D_g'$ gives the heating rate due to collisions with the background gas and $D_E'$ the heating rate due to electric field noise. }
% 	\label{plot.heatingrates}
% \end{figure}
% Here we can see how the two dominant heating sources scale with frequency; the gas, evidently, dominates at high frequencies; the electric field at lower. 
\subsection{APPENDIX 4: An Analytic Solution for A Noise With Gaussian Spectrum\label{gaussapp}}
% \footnotetext[1]{$\eta$'s proportionality to $\lambda$ may not be linear, and depends upon the size and geometry of the object. A calculation of $\eta$ for a nanosphere of the type under consideration here is included in the supplementary material.}

% (see figure \ref{plot.heatingrates})[REFERENCE TO BE FIXED]. Combining this with a bath possessing the correlation function given above, centred about $\nu_0=10^5$ Hz, with a width determined by $\gamma=10^4$ Hz and a coupling strength of $\eta=2.3 \times 10^{14}$ Hz, if we probe the system for a time $t=10^{-2}$ s (which guarantees a precision $1/t=100 $ Hz, a hundred times smaller than $\gamma$) we would recover information about that correlation function via the way in which the heating rate of the sphere would vary with the mechanical frequency, as seen in figure \ref{plot.CorrelationCompare}.

% \begin{figure}
%     \centering
%     \includegraphics[width=.45\textwidth]{./Plots/plotcompare3.pdf}
%     \caption{(Colour online). The heating rate of the sphere varies as a function of the mechanical frequency, showing a peak around the resonant frequency $\nu_0=10^5$ Hz of the collapsing field. The correlations function $\tilde{C}(\omega)$ is shown (and amplified by a factor of 1000) for comparison (note that this is not a heating rate, and the y axis label does not apply to this function). }
%     \label{plot.CorrelationCompare}
% \end{figure}

In this appendix, we wish to analyze explicitly the protocol suggested in the main text for the detection of non-Markovian noises in the case of a bath/noise with a Gaussian spectrum of the form 
\begin{equation}
\tilde{C}(\nu)= \eta e^{-\frac{(\nu_0-\nu)^2}{2\gamma^2}}\label{eq.SIGAU}
\end{equation}
in which $\eta$   gives the strength of the coupling of the noise field to the nanosphere, $\nu_0$ gives the resonant frequency of the collapse noise, and $\gamma$ gives the width of the spectrum. 
Inserting $\tilde{C}(\nu)$ from Eq. \eqref{eq.SIGAU} in Eq. (13) from the main text one gets:

\begin{align}
\langle n\rangle_t = \langle n\rangle_0 +D_p't+\mathfrak{N}_{t}, \label{eq.gaussian corr}
\end{align}
where
\begin{align}
\mathfrak{N}_{t}&=\frac{\eta}{2\pi m\omega_m}\int_{-\infty}^{+\infty}d\nu\,e^{-\frac{(\nu_{0}-\nu)^{2}}{2\gamma^{2}}}\frac{\sin^{2}[(\omega_{m}-\nu)t/2]}{(\omega_{m}-\nu)^{2}}=\frac{\eta}{2\pi m\omega_m}\int_{-\infty}^{+\infty}dy\,e^{-\frac{[y-(\omega_{m}-\nu_{0})]^{2}}{2\gamma^{2}}}\frac{\sin^{2}(yt/2)}{y^{2}}.
\end{align}
After rewriting the last term in the integral as
\begin{equation}
\frac{\sin(yt/2)}{y}=\frac{1}{4}\int_{-t}^{+t}e^{i\frac{y}{2}s}\,ds,
\end{equation}
we can perform the integration on $y$, which gives
\begin{align}
\mathfrak{N}_{t}&=\frac{\eta\gamma}{16 \sqrt{2 \pi} m \omega_m}\int_{-t}^{+t}\!\!ds\int_{-t}^{+t}\!\!ds'e^{-\frac{\gamma^{2}}{8}(s+s')^{2}-\frac{i}{2}(\ensuremath{\nu_{0}}-\omega)(s+s')}.
\end{align}
Making the change of variables $x_1=s-s'$ and $x_2=s+s'$, using the rule

\begin{equation}
\int_{-t}^{+t}\,ds\int_{-t}^{+t}\,ds'f\left(s,s'\right)=\frac{1}{2}\int_{0}^{2t}dx_2\int_{-\left(2t-x_2\right)}^{+\left(2t-x_2\right)}dx_1\,\left[f\left(x_1,x_2\right)+f\left(x_1,-x_2\right)\right]
\end{equation}
and introducing $z=x_2\gamma/2$, we get 
\begin{align}\label{NtGau}
\mathfrak{N}_{t}=\frac{\eta}{2\gamma m \omega_m}\sqrt{\frac{1}{2 \pi}}\int_{0}^{t\gamma}dz\left(\gamma t-z\right)e^{-\frac{z^{2}}{2}}\cos\left[\frac{(\ensuremath{\nu_{0}}-\omega_{m})}{\gamma}z\right].
\end{align}
Keeping in mind that the bath correlation is centered in $\nu_0$ with a width $\gamma$ while the $\omega_m$ describes the frequency we are probing with an error given by $1/t$, it is instructive to consider two limit cases:

1. $\gamma\ll1/t$ i.e. the error in probing the noise spectrum is much larger than its width. Then we can approximate the Gaussian inside Eq. \eqref{NtGau} as $e^{-\frac{z^{2}}{2}}\simeq1$ and get
\begin{align}
\mathfrak{N}_{t}\simeq\frac{\eta\gamma}{2m\omega_m}\sqrt{\frac{1}{2\pi}}\frac{1-\cos\left[(\ensuremath{\nu_{0}}-\omega_{m})t\right]}{(\ensuremath{\nu_{0}}-\omega_{m})^{2}}
\end{align}

The above formula gives the largest contributions when $|\ensuremath{\nu_{0}}-\omega_{m}|\leq1/t$ (this is reasonable: if the distance between $\ensuremath{\nu_{0}}$ and $\omega_{m}$ is larger than the probe precision $1/t$, the probe cannot “see” $\ensuremath{\nu_{0}}$) in which case we can further approximate $\mathfrak{N}_{t}\simeq\sqrt{\frac{1}{2\pi}}\frac{\eta\gamma}{4m\omega_m}t^{2}$. However, all this also holds whether $\ensuremath{|\nu_{0}}-\omega_{m}|\gg\gamma$ or $\ensuremath{|\nu_{0}}-\omega_{m}|\ll\gamma$, whereas a good probe requires that for $\ensuremath{|\nu_{0}}-\omega_{m}|\gg\gamma$ the above should be zero. As expected, if $\gamma \ll 1/t$ then the probe is not effective for the bath considered here. 

2. $\gamma\gg1/t$ i.e. the error in probing the noise spectrum is much smaller than its width. Then
\begin{align}
\mathfrak{N}_{t}&\simeq\frac{\eta}{2\gamma m \omega_m}\sqrt{\frac{1}{2\pi}}\gamma t\int_{0}^{t\gamma}ds\,e^{-\frac{s^{2}}{2}}\cos\left[\frac{(\ensuremath{\nu_{0}}-\omega_{m})}{\gamma}s\right]\\
&\simeq\frac{\eta}{2\gamma m \omega_m}\sqrt{\frac{1}{2\pi}}\gamma t\int_{0}^{\infty}ds\,e^{-\frac{s^{2}}{2}}\cos\left[\frac{(\ensuremath{\nu_{0}}-\omega_{m})}{\gamma}s\right]\nonumber\\
&=\frac{\eta t}{4 m \omega_m} e^{-\frac{(\ensuremath{\nu_{0}}-\omega_{m})^{2}}{2\gamma^{2}}},\nonumber
\end{align}
which is exactly what is expected by a good probe: the contributions are relevat only in the region $\omega_m\simeq[\nu_0-\gamma,\nu_0+\gamma]$ where the bath correlation is non zero.

\subsection{APPENDIX 5: On the ringing seen for low fidelity reconstructions\label{sec.Ringing}}

In figure 1 in the main text, we observe that there is a characteristic `ring-down' effect, seen only on the purple line ($t=10^{-4}$ s), and at frequencies $\omega_m>10^5$ Hz. We stated in the main text that this constitutes a feature of the spectrometer -- that it can be used to distinguish between two situations: those in which the experimenter has missed features of the noise spectrum owing to an inadequate measurement time; and those in which the measurement times has been sufficient to recover all salient features up to the frequency explored. Here we will explain this claim. 
\paragraph{}
To begin, we consider Eq. (13) from the main text:
\[
\langle\hat{n}\rangle_{t}=\langle\hat{n}\rangle_{0}+\frac{1}{2\pi m\omega_{m}}\int_{-\infty}^{\infty}d\nu\,\tilde{C}(\nu)\frac{\sin^{2}[(\omega_{m}-\nu)t/2]}{(\omega_{m}-\nu)^{2}}.
\]
For the sake of illustrating the effect which we're interested in, let's suppose that $\tilde{C}(\nu)$ presents a sharp peak with a width
$\ll1/t$ around a certain point $\nu_{0}$, and that it is close to zero elsewhere. In this case, the above equation would approximate to  
\begin{equation}
\langle\hat{n}\rangle_{t}\simeq\langle\hat{n}\rangle_{0}+\frac{A}{2\pi m\omega_{m}}\frac{\sin^{2}[(\omega_{m}-\nu_{0})t/2]}{(\omega_{m}-\nu_{0})^{2}}\label{nt_appro}
\end{equation}
with $A:=\int_{-\infty}^{\infty}d\nu\,\tilde{C}(\nu)<\infty$ for
any realistic noise. So, as we scan through $\omega_m$, we will see the oscillating on the $\sin^2$ term -- exactly what we see on the purple line in figure 1. 

How should we understand this behaviour? Essentially, when a feature of $\tilde{C}(\nu)$ has a width of more then $\sim1/t$, then the $\sin^2[(\omega_m-\nu)t/2]/(\omega_m-\nu)^2$ term acts as a filter, and probes $\tilde{C}(\nu)$ by varying $\omega_m$. If, however, feature of the noise spectrum is sharper than $\sim1/t$, then this relationship becomes reversed: as $\omega_m$ is varied, it is as if the noise spectrum were sampling the ${\rm sinc}^2$ term, and recovering the ringing. 
% Clearly, the higher the
% distance between $\omega_{m}$ and the center of the peak in the spectrum
% $\nu_{0}$, the smaller it is the effect due to the quadratic term
% in the denominator, a feature which can be also seen in Fig. (YYY):
% we observe the oscillation from $\omega_{m}\simeq10^{5}$ Hz for $t=10^{-4}$
% s and then their intensity is more weak. 

% Basically, what is happening here is that since the spectrum we are
% measuring is much sharper than the sinc squared function, they roles
% are reversed: $\tilde{C}(\nu)$ acts as a filter and what we are determining
% is the shape of the squared sinc function. 

This reasoning can be generalized also to the case when there are multiple
peaks at the frequencies $\nu_{0}$,$\nu_{1}$...$\nu_{N}$, in which case Eq. (\ref{nt_appro}) becomes:
\begin{equation}
\langle\hat{n}\rangle_{t}\simeq\langle\hat{n}\rangle_{0}+\frac{A}{2\pi m\omega_{m}}\sum_{k=0}^{N}\frac{\sin^{2}[(\omega_{m}-\nu_{k})t/2]}{(\omega_{m}-\nu_{k})^{2}}.\label{nt_appro-1}
\end{equation}
We now want to consider the case when the spectrum has peaks
at the frequencies $\nu_{0}$,$\nu_{1}$...$\nu_{N}$, as in \eqref{nt_appro-1}, but in which the spectrum is not zero in the other points. Then we can write 
\[
\tilde{C}(\nu)=\tilde{C}_{s}(\nu)+\tilde{C}_{p}(\nu)
\]
where $\tilde{C}_{s}(\nu)$ contains the smooth part of the spectrum
% (changing much slower than $1/t$) 
and $\tilde{C}_{p}(\nu)$ the peaks. Then eq. (13) can be approximated as:
\begin{equation}
\langle\hat{n}\rangle_{t}\simeq\langle\hat{n}\rangle_{0}+\frac{1}{2\pi m\omega_{m}}\left[\tilde{C}_{s}(\omega_{m})\frac{\pi}{2}t+A_{p}\sum_{k=0}^{N}\frac{\sin^{2}[(\omega_{m}-\nu_{k})t/2]}{(\omega_{m}-\nu_{k})^{2}}\right]\label{nt_appro-2}
\end{equation}
where $A_{p}:=\int_{-\infty}^{\infty}d\nu\,\tilde{C}(\nu)$. Using
Eq. (14) from the main text to reconstruct the spectrum we would get:
\begin{equation}
\tilde{C}(\omega_{m})=\tilde{C}_{s}(\omega_{m})+\frac{2A_{p}}{\pi t}\sum_{k=0}^{N}\frac{\sin^{2}[(\omega_{m}-\nu_{k})t/2]}{(\omega_{m}-\nu_{k})^{2}}.\label{C final2}
\end{equation}
For very small times $t$, the second term goes to zero and, as expected
we can only reconstruct the smooth part of the spectrum $\tilde{C}_{s}(\omega_{m})$.
In the opposite limit, when $t$ is so large to allow to resolve the
peaks, we expect to perfectly reconstruct  the spectrum $\tilde{C}(\omega_{m})=\tilde{C}_{s}(\omega_{m})+\tilde{C}_{p}(\omega_{m})$.
If the time is not large enough to resolve the peaks but also not
so small to make the second term in Eq. (\ref{C final2}) completely
negligible, we will see some ringing in the spectrum which will clearly
tell us that more accuracy is required. This ringing we will see can be identified by its frequency -- it will be exactly that which is fixed by the experimenter's choice of $t$ and the ${\rm sinc}^2$ function -- in this way, the experimenter can be certain that he or she is not mistaking a genuine feature of the noise spectrum for a ringing which is owed to too short a measurement time. 

The example worked through here is exactly the  type of structure which is depicted in figure 1 -- a sum of  Gaussian peaks with different weights. For the red and green lines, all of the peaks being probed are of a width larger than $1/t$, whereas for the purple line, some peaks are sharper than $1/t$, which creates the ring-down behaviour.

\bibliography{library,ManualLibrary}
\end{document}